\documentclass[
    aps,
    prapplied,
    10pt,
    longbibliography,
    amsfonts,
    amssymb,
    amsmath,
    superscriptaddress,
    twocolumn,
    floatfix
]{revtex4-2}
\pdfoutput=1

\usepackage{cprotect}
\usepackage{csquotes}
\usepackage{dsfont}
\usepackage{environ}
\usepackage[a4paper, total={510pt, 678pt}]{geometry}
\usepackage{graphicx}
\usepackage[colorlinks]{hyperref}
\usepackage{physics}
\usepackage{placeins}
\usepackage{siunitx}
\usepackage{subcaption}
\usepackage{xcolor}
\usepackage{ragged2e}
\DeclareCaptionJustification{justified}{\justifying}
\definecolor{qmla_red}{HTML}{C73626}
\definecolor{qmla_blue}{HTML}{215CAF}
\hypersetup{
    colorlinks=true,
    linkcolor=qmla_red,
    citecolor=qmla_red,
    filecolor=qmla_blue,
    urlcolor=qmla_blue,
    pdftitle={Learning agent-based approach to the characterization of open quantum systems},
}
\newcommand{\cprim}[1]{\mathcal{H}\qty[#1]}
\newcommand{\dprim}[1]{\mathcal{D}\qty[#1]}
\NewEnviron{peq}[1]{ \begin{equation}\phantom{#1}\BODY #1\end{equation} }

\begin{document}


    \title{Learning agent-based approach to the characterization of open
    quantum systems}

    \author{ \firstname{Lorenzo} \surname{Fioroni} }
    \email{lorenzo.fioroni@epfl.ch}
    \thanks{Present address: Institute of Physics, École Polytechnique Fédérale
    de Lausanne (EPFL), 1015 Lausanne, Switzerland (CH)}
    \affiliation{Institute for Quantum Electronics, ETH Z\"urich, 8093 Z\"urich, Switzerland}

    \author{ \firstname{Ivan} \surname{Rojkov} }
    \affiliation{Institute for Quantum Electronics, ETH Z\"urich, 8093 Z\"urich, Switzerland}
    \affiliation{Quantum Center, ETH Zürich, 8093 Zürich, Switzerland}

    \author{ \firstname{Florentin} \surname{Reiter} }
    \affiliation{Institute for Quantum Electronics, ETH Z\"urich, 8093 Z\"urich, Switzerland}
    \affiliation{Quantum Center, ETH Zürich, 8093 Zürich, Switzerland}
    \affiliation{Fraunhofer Institute for Applied Solid State Physics IAF, Tullastra{\ss}e 72, 79108 Freiburg, Germany}


    \begin{abstract}
        Characterizing quantum processes is crucial for the execution of quantum algorithms on available quantum devices.
        A powerful framework for this purpose is the \emph{Quantum Model Learning Agent (QMLA)} which characterizes a given system by learning its Hamiltonian via adaptive generations of informative experiments and their validation against simulated models.
        Identifying the incoherent noise of a quantum device in addition to its coherent interactions is, however, as essential.
        Precise knowledge of such imperfections of a quantum device allows to devise strategies to mitigate detrimental effects, for example via quantum error correction. 
        We introduce the \emph{open} Quantum Model Learning Agent (oQMLA) framework to account for Markovian noise through the Liouvillian formalism.
        By simultaneously learning the Hamiltonian and jump operators, oQMLA independently captures both the coherent and incoherent dynamics of a system.
        The added complexity of open systems necessitates advanced algorithmic strategies.
        Among these, we implement regularization to steer the algorithm towards plausible models and an unbiased metric to evaluate the quality of the results.
        We validate our implementation in simulated scenarios of increasing complexity, demonstrating its robustness to hardware-induced measurement errors and its ability to characterize systems using only local operations.
        Additionally, we develop a scheme to interface oQMLA with a publicly available superconducting quantum computer, showcasing its practical utility.
        These advancements represent a significant step toward improving the performance of quantum hardware and contribute to the broader goal of advancing quantum technologies and their applications.
    \end{abstract}

    \maketitle


    \section{Introduction}
        Accurately characterizing quantum processes is pivotal for advancing quantum technologies.
        This involves modeling both the coherent interactions within a system and the incoherent processes arising from its coupling with an often unknown environment.
        Understanding these dissipative dynamics is especially critical for designing robust quantum devices, as it enables the development of fault-tolerant protocols, including error mitigation and correction in quantum computation~\cite{werninghaus_leakage_2021, Egger_adaptive_2014}, as well as the verification of quantum communication schemes~\cite{Lobino_memory_2009}. 
        A common way to study dissipative processes relies on the Choi-Jamiolkowski isomorphism~\cite{choi_completely_1975, jamiolkowski_linear_1972} to map an unknown evolution to a state in a higher-dimensional Hilbert space, subsequently reconstructed through complete quantum state tomography~\cite{mohseni_quantum-process_2008, gebhart_learning_2022}. 
        Despite its conceptual simplicity, this approach faces interpretability and scalability issues with larger systems due to the exponential growth of the number of parameters to be estimated and the abstract representation of the unknown environment.
        In the pursuit of scalable alternatives, several methods have been proposed over the years, seeking a balance between the complexity and the amount of information gathered about the process~\cite{nielsen_gate_2021, knill_randomized_2008, 
        Holzpfel2015, riofrio_experimental_2017, 
        ahmedGradientDescentQuantumProcess2023, kaufmannCharacterizationCoherentErrors2024}.
        A paradigmatic example is the \emph{randomized benchmarking} method, that estimates the average fidelity of a process by interleaving it with a sequence of random gates, and analyzing the system's final state~\cite{knill_randomized_2008, magesan_scalable_2011}. 
        Other methods leverage recent advancements in machine learning to provide approximate representation of quantum systems. 
        This includes techniques like \emph{Neural network quantum states}~\cite{torlai_neural-network_2018, carleo_solving_2017, carleo_constructing_2018} and \emph{Quantum generative adversarial networks}~\cite{ahmed_quantum_2021}.

        \begin{figure*}[htb]
        \centering
        \includegraphics{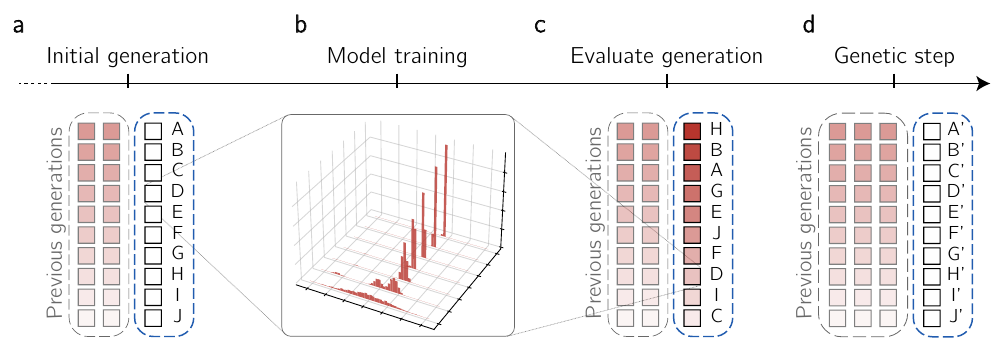}
        {\phantomsubcaption
        \label{fig: QMLA overview: initial generation}}
        {\phantomsubcaption
        \label{fig: QMLA overview: model training}}
        {\phantomsubcaption
        \label{fig: QMLA overview: evaluate generation}}
        {\phantomsubcaption
        \label{fig: QMLA overview: genetic step}}
        \caption{
            \emph{Schematic of a single step of the oQMLA algorithm} - \textbf{a)}~Independent models are held in generations. 
            Colored boxes illustrate models that have previously been trained and ranked, while the white ones represent explicitly defined models yet to be trained. 
            \textbf{b)}~Each one of the models in the current generation is trained via a Bayesian inference sub-routine until its optimal parametrization is found. 
            \textbf{c)}~The trained models are tested against experimental evidence and ranked by their performance. 
            \textbf{d)}~The genetic step uses the information about the performance of the trained models to propose a new generation of tentative models.
        }
        \label{fig: QMLA overview}
        \end{figure*}

        Algorithms based on the recently introduced \emph{quantum model learning agent} (QMLA) paradigm represent yet another example within the framework of process tomography methods employing machine learning techniques~\cite{gentile_learning_2021, flynn_quantum_2022}.
        QMLA aims to find the most likely Hamiltonian model constructed as a linear combination of Hermitian operators that describes the evolution of a closed system. 
        To achieve this, the algorithm employs a double-nested optimization process. 
        First, it considers several possible models for the evolution, estimates their parameters, and tests them against experimental evidence.  
        Subsequently, their performance is assessed, and the results are employed to explore the model space without imposing constraints on the structure of the true Hamiltonian.
        Upon convergence, it outputs an approximation of the system's model, consisting of a collection of operators describing its evolution. 
        Results in this form are more informative compared to those of benchmarking techniques, as they provide complete information about the evolution of the system, while still requiring fewer measurements compared to full tomographic methods~\cite{flynn_quantum_2022}. 
        Moreover, unlike full tomographic methods that aim to estimate a matrix representation of a given process, QMLA's goal is to identify the operators composing such a matrix. 
        Therefore, it yields interpretable information about the model, providing details about the couplings present within the system.
        Apart from a recent work that implements a similar technique for characterizing the dynamics of open systems, albeit focused on distinct use cases~\cite{wallaceLearningDynamicsMarkovian2024}, QMLA's primary assumption is that the considered system is closed.
        Dissipative processes can be incorporated by studying the evolution of the quantum state representing both system and environment together~\cite{gentile_learning_2021}. 
        However, this approach is not always feasible as it relies on the hypothesis that the environment can be faithfully modeled and treated analytically. 

        In this article, we revise the QMLA method to the case of open quantum systems by modeling their Markovian dynamics using the Lindblad master equation. We name the resulting framework the \emph{open} Quantum Model Learning Agent (oQMLA). As illustrated in Fig.~\ref{fig: QMLA overview}, oQMLA implements the two optimization layers of QMLA via Bayesian inference and a genetic algorithm, respectively.
        The former allows estimating the optimal parameters of each Liouvillian model considered, while the latter is responsible for generating potential models to be tested.
        By leveraging the speed-up enabled by this choice, our approach can characterize the dynamics of a system in the presence of noise. 
        The characterization result provides details about the Hamiltonian describing its coherent evolution, as well as the jump operators describing the interaction with the environment. 
        We identify the challenges that arise from the application of QMLA to open systems and propose solutions to address them. 
        In particular, we advance the use of a regularization technique to bias the algorithm towards plausible models contrasting the growth of the model space and adopt the \emph{root mean squared error} (RMSE) as a figure of merit to provide an unbiased evaluation of the models' quality.
        The performance of our method is benchmarked in a series of increasingly complicated classical simulations, demonstrating its resilience towards some limitations of realistic noisy devices such as readout errors and the availability of a restricted set of operations. 
        Finally, we interface our algorithm with a gate-based quantum computer utilizing superconducting circuits and characterize the noise impacting the evolution of a two-qubit controlled-NOT gate on real hardware.

        Our findings suggest that the approach presented here could be employed to provide valuable information about the evolution of a system, allowing for the identification of the most prominent error sources. 
        The analysis of oQMLA's output could thus facilitate the design of error mitigation techniques and allow for better calibration of hardware devices~\cite{werninghaus_leakage_2021, temme_error_2017, pastori_characterization_2022, werninghaus_leakage_2021, Egger_adaptive_2014, Lobino_memory_2009}.


    \section{Quantum model learning agent}

        The QMLA method aims to find the most likely model for the evolution of a system, from within a predefined collection of models. 
        Previous works focused on the study of coherent evolutions~\cite{wang_experimental_2017, santagati_magnetic-field_2019, gentile_learning_2021, flynn_quantum_2022}, where tentative Hamiltonian models were defined as weighted sums of operators (referred to as \emph{primitives}) in a set $\mathcal{S} = \qty{h_i \mid h_i \text{ is Hermitian} }$:
        \begin{equation}
            H = \sum_i \alpha_i h_i; \quad \alpha_i \in \mathbb{R} \; \forall i.
        \end{equation}
        By considering multiple models in a tree-like structure, QMLA \emph{trains} all the models in a branch concurrently; that is, it finds the most likely multiplicative factors $\alpha_i$ in front of each one of the model's primitives $h_i$. 
        The performance of the trained models is subsequently tested against experimental evidence. 
        Finally, an \emph{exploration strategy} determines how the tree should evolve, either defining a new collection of models to train and test, or terminating.
        
        The remainder of this section is dedicated to providing a detailed introduction to the components constituting our characterization framework for open systems: the oQMLA algorithm. 
        These are equivalently sketched in Fig.~\ref{fig: QMLA overview}. 
        Although some of the elements align with their closed-system counterparts, notable differences are present and will be addressed in subsequent sections.
        
        \subsection{Model}
        \label{sec: model}
        Despite being effective in identifying the Hamiltonian model that best describes the dynamics of a system, the QMLA framework cannot be directly applied to the analysis of an open system. 
        In order to correctly capture the interactions with the environment, in fact, it requires including the latter in the system's description.
        We obtain an effective representation of the dynamics restricted to the system only by modeling the evolution via the diagonal form of the Lindblad master equation~\cite{breuer_theory_2007, manzano_short_2020}
        \begin{peq}{.}
            \pdv{\rho}{t} = \sum_i \alpha_i \cprim{h_i}\qty(\rho) + \sum_k \Gamma_k \dprim{L_k}\qty(\rho)
            \label{eq: Lindblad}
        \end{peq}

        The functionals in Eq.~\eqref{eq: Lindblad} describe the coherent and dissipative evolutions of the system respectively, and are defined as 
        \begin{align}
            \phantom{;}\cprim{h} &= -i\comm{h}{\rho};\label{eq: coherent functional} \\
            \phantom{.}\dprim{L} &= L\rho L^\dagger - \frac{1}{2}\acomm{L^\dagger L}{\rho}.\label{eq: dissipative functional}
        \end{align}
        
        The Lindblad in Eq.\eqref{eq: Lindblad} provides an effective description of the reduced dynamics of an open quantum system under a set of well-established approximations. 
        First, the coupling to the environment is assumed to be weak (Born approximation).
        Second, the environment's correlations must decay on timescales much shorter than the system's evolution, justifying the neglect of memory effects (Markov approximation). 
        Finally, the rotating-wave approximation eliminates the terms that oscillate rapidly in time. 
        The \emph{oQMLA} framework introduced in this work is applicable to systems where these conditions are satisfied. 
        This includes, for example, superconducting qubits operated in suitable regimes~\cite{Samach2022, Gambetta2008, Boissonneault2009}, trapped-ion systems~\cite{brownnutt_2015,nakav_2023}, and Rydberg atom arrays~\cite{Lee2019}.
        
        The choice of using Eq.~\eqref{eq: Lindblad} allows us to define models as weighted sums of elements from a set $\mathcal{S}$. Elements of this set are the functionals defined in equations~\eqref{eq: coherent functional} and~\eqref{eq: dissipative functional}, i.e. ${\mathcal{S} = \qty{\cprim{h_i} \mid h_i \text{ is Hermitian}} \bigcup \qty{\dprim{L_j}}}$. 
        Note that using the diagonal master equation limits the expressive power of the models. 
        In fact, to enable a set $\mathcal{S}$ to generate any arbitrary dynamics, it must include the dissipators $\dprim{L_k}$ for all continuous combinations of a complete set of operators.
        The cardinality of the set $\mathcal{S}$ thus influences the probability of achieving a more accurate approximation of the true model. 
        We increase this probability by considering the set of jump operators to be an overcomplete set of operators.
        
        In analogy to the standard QMLA, we define tentative models as weighted combinations of elements in~$\mathcal{S}$.
        
        \subsection{Training}
        \label{sec: training}
        After defining the tentative models, their parameters need to be optimized. 
        oQMLA does not impose a specific optimization algorithm. 
        In this work, we find the most likely parametrization of the model using Bayesian inference~\cite{santagati_magnetic-field_2019, wang_experimental_2017, granade_robust_2012, wiebe_hamiltonian_2014}:
        
        Considering a model $m\qty(\vb{x})$ which depends on a set of (fixed) primitives and a vector of parameters $\vb{x}$, we assume a prior distribution $P_0\qty(\vb{x})$ for the parametrization and evolve it according to the Bayes rule
        \begin{peq}{,}
            \Pr(\vb{x}) \leftarrow \Pr(\vb{x}|d) \propto \Pr(d|\vb{x}) \Pr(\vb{x})
        \end{peq}
        where $d$ is the outcome of an experiment performed on the system and $\Pr(d|\vb{x})$ is the probability of obtaining such an outcome, evaluated through a classical simulation.
        The update is performed using the sequential Monte Carlo method~\cite{doucet_sequential_2001}, which approximates the probability density function by utilizing a finite set of samples, referred to as \emph{particles}, as its support. 
        Each particle is assigned a weight to represent its significance; updating such weights suffices to update the distribution of the parameters. 
        At any given time, the best estimate of the real parametrization is provided by $\mathbb{E}\qty[\vb{x}] \approx \sum_k w_k \vb{x}_k = \tilde{\vb{x}}$, where $w_k$ is the weight associated with the particle $\vb{x}_k$. 
        Similarly, the mean dispersion of the distribution $\sigma = \sqrt{\sum_k w_k \qty|\vb{x}_k - \tilde{\vb{x}}|^2}$ serves as an indication of the algorithm's convergence.
        
        Notably, the performed experiments can be tailored to the specific characteristics of the experimental apparatus.
        We identify every experiment by a tuple $\qty(\rho_0,\,t,\,\mathcal{U})$, denoting the preparation of a state $\rho_0$, its evolution for a time $t$ according to the underlying model followed by the application of the unitary $\mathcal{U}$ and a measurement in the computational basis.
        For the simulation, we assume that any pure initial state $\rho_0$ can be prepared and any possible unitary $\mathcal{U}$ can be applied.
        In Section~\ref{sec: local operations}, we will examine how the learning process is influenced when we relax this assumption. 
        The choice of the evolution time is based on the \emph{particle guess heuristic}~\cite{santagati_magnetic-field_2019, wang_experimental_2017, gentile_learning_2021, wiebe_hamiltonian_2014, granade_structured_2017}, which modulates $t$ depending on the uncertainty on the rates:
        \begin{peq}{.}
            t = \frac{1}{\sigma}
            \label{eq: particle guess heuristic}
        \end{peq}
        
        \begin{figure*}[ht!]
        \centering
        \includegraphics{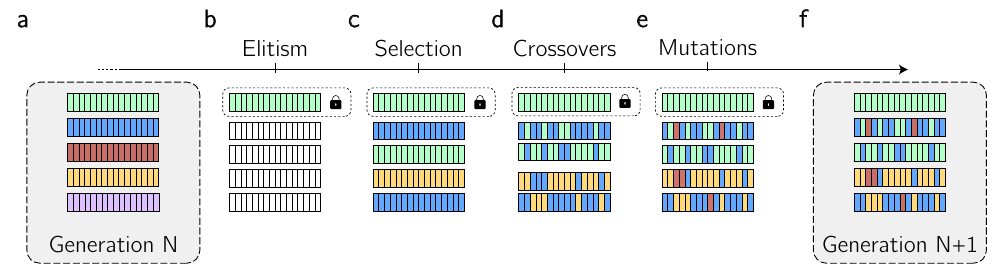}
        {\phantomsubcaption\label{fig: genetic step overview: initial}}
        {\phantomsubcaption\label{fig: genetic step overview: elitism}}
        {\phantomsubcaption\label{fig: genetic step overview: selection}}
        {\phantomsubcaption\label{fig: genetic step overview: crossovers}}
        {\phantomsubcaption\label{fig: genetic step overview: mutations}}
        {\phantomsubcaption\label{fig: genetic step overview: final}}
        \caption{
            \emph{Schematic of the application of the genetic step} - \textbf{a)}~The models in the parent generation are trained and ranked by their performance. 
            \textbf{b)}~Elitism is enforced: the best model survives the genetic step and is preserved across the subsequent operations. 
            \textbf{c)}~Models from the parent generation are randomly selected according to their fitness value. 
            \textbf{d)}~Application of crossovers between the selected models. 
            The elitist models do not undergo crossovers. 
            \textbf{e)}~Random mutations are applied to all the models except the elitist ones. 
            \textbf{f)}~Models of the offspring generation. 
        }
        \label{fig: genetic step overview}
        \end{figure*}
        
        \subsection{Testing}
        Once the models of a generation are trained, their performance is assessed by comparing them against experimental data. 
        Various methods have been previously proposed to evaluate the quality of models trained via Bayesian inference, such as \emph{inverse log-likelihood}, \emph{Bayes factor points} or  \emph{residuals}~\cite{flynn_quantum_2022}.
        A typical choice of metric is the \emph{Bayes factors} (BFs)~\cite{gentile_learning_2021}, identifying which one, between two models, is more likely to be correct. 
        The Bayes factor between models $m_1$ and $m_2$ is defined as the exponential of the ratio between the likelihoods, denoted as $\mathcal{L}\qty(m)$, of observing the outcomes in the test set under the assumptions of model $m_1$ and $m_2$ respectively: $\text{BF}\qty(m_1,\, m_2) = \exp(\mathcal{L}\qty(m_1) / \mathcal{L}\qty(m_2))$.
        
        Bayes factors offer a means to rank models within a given generation. 
        However, since they only provide rankings, they are limited in their ability to quantify the absolute quality of these models.
        A more suitable figure of merit for the model's performance should therefore be a function $f\in\qty[0;\,\infty]$ that grows monotonically with improvements in model quality and depends exclusively on the model $m$ it is evaluated on. 
        We refer to $f$ as the \emph{fitness function}. 
        
        A function satisfying these conditions is one that solely depends on fitting the statistics of the outcomes, effectively removing the dependency on the specific shape of the model and hence representing an unbiased estimator of its quality. 
        Each experiment $e_i$ in a test set $\mathcal{T}$ is associated with the vector of the outcome probabilities $\vb{p}_i$. 
        This can be estimated from a set of measurement outcomes, or via more efficient techniques like \emph{Classical Shadows}~\cite{huang_predicting_2020}
        For a given model $m$, we classically simulate the probability vector $\vb{q}_i$ that one would expect to observe if $m$ was correct.
        A measure of the agreement between estimated and true probabilities is provided by the \emph{root mean squared error} (RMSE) defined as
        \begin{peq}{,}
             \sqrt{\frac{1}{\qty(2^n-1)\qty|\mathcal{T}|}\sum_i \sum_{j=0}^{2^n-2}\qty|q_{i,j} - p_{i,j}|^2}
        \end{peq}
        where $q_{i,j}$ and $p_{i,j}$ are the $j$-th components of the $\vb{q}_i$ and $\vb{p}_i$ vectors respectively. 
        The RMSE indicates the average difference between the estimated probabilities and the true ones, with smaller values indicating better agreement. 
        Defining the fitness function as 
        \begin{peq}{,}
            f\qty(m) = \frac{1}{\text{RMSE}}
            \label{eq: fitness}
        \end{peq}
        ensures its monotonic growth with the performance of the model. 
        Note that in computing the RMSE we ignored the last component of the probability vectors that is constrained by their normalization and including it would obscure its intuitive meaning.
        
        \subsection{Reproduction}
        \label{sec: reproduction}
        
        An \emph{exploration strategy} determines how new models are generated from the previously tested ones.
        As said above, the space of representable models is determined by the choice of $\mathcal{S}$, and including several jump operators in it increases the chances of finding a good approximation of the true model, at the cost of the search complexity.
        As an example, considering a 2-qubit system with $10$ primitives to describe each qubit ($4$ coherent ones and $6$ jump operators) and taking into account all the possible tensor products, the cardinality of $\mathcal{S}$ amounts to $50$ independent operators. 
        Since the models are encoded in $\qty|\mathcal{S}|$\mbox{-}bits strings, the model space has a dimension of $2^{50}$. 
        In this work, we utilize a genetic algorithm for this task~\cite{gentile_learning_2021, flynn_quantum_2022}.  
        First, each model is encoded in a $\qty|\mathcal{S}|$\mbox{-}bits string, referred to as  \emph{chromosome}. 
        A $1$ ($0$) in position $i$ signals the presence (absence) in the model of the $i$\mbox{-}th primitive defined in $\mathcal{S}$. 
        Once all models in a branch have been evaluated, the genetic algorithm executes the \emph{genetic step} (GS) according to their scores~\cite{goldberg_genetic_1988}. 
        This consists of a sequence of four operations on the bit\mbox{-}strings to output an offspring achieving on average a higher score. 
        Fig.~\ref{fig: genetic step overview} shows a schematic representation of the operations of the genetic step. 
        
        First, we enforce \emph{elitism} by copying the individual achieving the highest value of the fitness in generation $N$ to generation $N+1$. 
        Elitism ensures that the quality of the models found by oQMLA can only increase with the number of generations, and consequently, it is a necessary condition for the convergence of the algorithm~\cite{rudolph_convergence_1994, iosifescu_finite_2014}. 
        Pairs of \emph{parents} from generation $N$ are then selected via the \emph{roulette wheel selection} method~\cite{blickle_comparison_1996}. 
        That is, each model is selected to be parent with a probability $p\qty(m) \propto g\qty(f\qty(m))$,
        where $g$ is a monotonic function to account for the scaling of~$f$. 
        To capture small improvements of $f$ from the very early stage of the algorithm we set $g$ to be an exponential function.
        The \emph{crossover} operation intermixes the chromosomes of the parent models to produce those of the offspring. 
        Different crossover schemes have been used in genetic algorithms~\cite{davis_handbook_1991}, and one- and two-fold crossovers have also been tested within the QMLA framework~\cite{gentile_learning_2021, flynn_quantum_2022}. 
        Since no information is encoded in the distance between two genes of the chromosome, we employ uniform crossovers, meaning that each of the genes of the children models is inherited by one of the two parents according to some fixed probability $p$.
        A higher value of $p$ results in a more pronounced mixing of the genes, allowing for a faster exploration of the model space. 
        However, it produces offspring models that resemble their ancestors less.
        Finally, random mutations ensure that the entire space of models can be explored regardless of the initial conditions~\cite{davis_handbook_1991}. 
        Each bit in the chromosome gets its value swapped according to fixed probabilities $P_{0\to 1}$ and $P_{1\to 0}$. 
        By choosing a target number of primitives $T$ and setting 
        \begin{peq}{,}
            P_{0\to 1} = \frac{T}{2^{\qty|\mathcal{S}|}-T} P_{1\to 0}
        \end{peq}
        we can use the mutations as a regularization technique, biasing the model search towards models that we regard as being physically plausible. 
        Indeed, setting $P_{0\to 1} = P_{1\to 0}$ and assuming that only mutations are applied, the models would evolve towards chromosomes with, on average $2^{\qty|\mathcal{S}|-1}$ different primitives, excessive for a physically meaningful system and the classical simulator. 
        It is important to stress that this regularization technique does not impose any structure on the model. 
        Due to the presence of the selection and crossovers operations, oQMLA can still find models of any number of primitives.


    \begin{figure*}[htb!]
        \centering
        \includegraphics{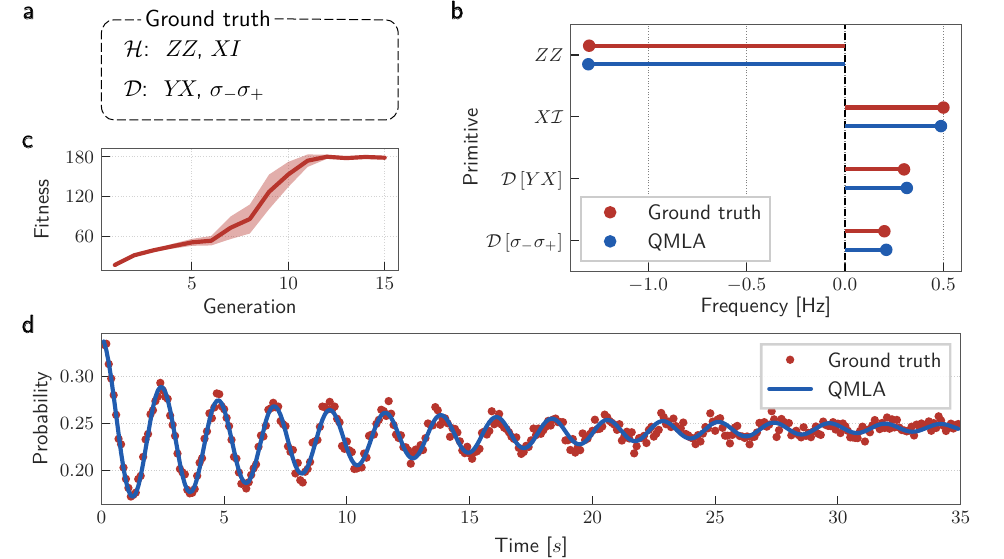}
        {\phantomsubcaption\label{fig: unlimited resources - model}}
        {\phantomsubcaption\label{fig: unlimited resources - lollipop}}
        {\phantomsubcaption\label{fig: unlimited resources - convergence}}
        {\phantomsubcaption\label{fig: unlimited resources - evolution}}
        \caption{
            \emph{Simulation of a two-qubit evolution} - We simulate an arbitrary evolution of a two-qubit open system and use oQMLA to learn it. 
            \textbf{a)}~ Primitives defining the true model. 
            \textbf{b)}~Comparison between oQMLA's output and the true model, both in terms of the primitives and the respective parameters.  
            \textbf{c)}~Evolution of the mean fitness value, computed on the best model of $5$ independent executions. 
            A narrow distribution signals resilience towards different initial conditions.  
            \textbf{d)}~Comparison between the dynamics generated by oQMLA's output and the true one, in a random experiment.
        }
        \label{fig: unlimited resources}
    \end{figure*}
        
    \section{Test cases}
        \label{sec: benchmarking}
        
        We evaluate the effectiveness of our implementation of the oQMLA algorithm by learning the evolution of simulated systems.
        The first tests are finalized at characterizing single- and two-qubit systems with arbitrary Hamiltonians $H$ and jump operators $L_k\in\mathcal{S}$.
        Subsequently, we analyze the capability of the algorithm to overcome typical challenges that stem from its application to real hardware devices.
        Specifically, we simulate the presence of measurement errors, allow oQMLA to perform local operations only, and simulate a system with jump operators not included in $\mathcal{S}$. In the latter case, oQMLA cannot find the true model of the system, but only an approximation which given a large enough set of primitives can reach an arbitrary precision.
        
        In the rest of this section, we focus on two-qubit systems, as they offer a vast enough model space to make the speed-up given by the genetic algorithm evident.
        In fact, even considering a comprehensive set of available primitives, it is likely that the overhead given by the use of the genetic algorithm on a single-qubit system outweighs its benefits. 
        Moreover, we argue that the most prominent noise sources on physical systems are single- and two-qubit processes. 
        oQMLA can be thus bootstrapped to learn such sources in pairs. 
        
        For the upcoming simulations, the set of available primitives $\mathcal{S}$ has been chosen to include all the Pauli operators and their tensor products for both the coherent and dissipative parts. For the latter, $\mathcal{S}$ also includes the raising and lowering operators on each qubit.
        
        \subsection{Two-qubit open system}
        We begin by examining the evolution of two qubits that interact with each other and the environment. 
        At this stage, we assume that the availability of resources is not a limiting factor, and thus we generate a highly comprehensive dataset where two qubits evolve from time \num{0} to \SI{35}{\second} in steps of \SI{e-3}{\second}, according to an arbitrary model
        \begin{equation*}
            \phantom{,}0.5\cprim{XI} + 1.3\cprim{Z Z} + 0.2\dprim{\sigma_-  \sigma_+} + 0.3\dprim{Y X},
        \end{equation*}
        where we denoted with $AB$ the tensor product operator $A\otimes B$ and with $\sigma_+$ and $\sigma_-$ the raising and lowering operators for the qubit, respectively.
        In total, the extremely fine sampling together with the choice of the final time sums up to $\mathcal{O}\qty(10^4)$ different experiments. 
        Although these values may seem excessive, oQMLA only uses a subset of the measured data: each model is trained on a maximum of $300$ experiments chosen accordingly to the particle-guess heuristic introduced in Eq.~\eqref{eq: particle guess heuristic}. 
        Moreover, each experiment is repeated only $50$ times, providing the algorithm with little information about the outcome statistics. 
        Nonetheless, opting for an extensive collection of training points allows to test the algorithm without assessing its limitations just yet. 
        Additionally, it allows to use the same dataset for multiple independent runs of the algorithm with negligible probability of training on the same outcomes. 
        The plot in Fig.~\ref{fig: unlimited resources - convergence} shows the mean value of the fitness function over five independent simulations. 
        Its value steadily increases and converges to values of approximately \num{180} after $12$ generations.
        Owing to the choice of the fitness function in Eq.~\eqref{eq: fitness}, we can relate these scores to the mean prediction error of the model. 
        Values of the fitness function greater than \num{180} correspond to a mean error in prediction below $6\text{\textperthousand}$, mainly due to statistical noise in the outcomes of the training set and the application of the Monte Carlo estimator. 
        The accordance between the predictions and the true data is visible in Fig.~\ref{fig: unlimited resources - evolution}, which displays the evolution of a randomly selected experiment as predicted by oQMLA, superimposed with the true one. 
        Finally, Fig.~\ref{fig: unlimited resources - lollipop} displays a comparison between the best model found and the true one. We see that the algorithm learned the correct primitives as well as an accurate parametrization. 
        Together, the plots in Fig.~\ref{fig: unlimited resources} show that, when the available resources are not limited, oQMLA can find the model describing the evolution of an interacting open two-qubit system in a few generations only.
        In particular, before the true model of the exemplified system has been found, oQMLA tested up to $300$ of the $2^{50}$ total models.
        
        \subsection{Local operations}
        \label{sec: local operations}
        The previous simulations assumed that the initial state $\ket{\psi}$ can be accurately prepared and that measurements in arbitrary bases can be mapped to measurements in the computational basis without introducing any error. 
        However, when applying the algorithm to real quantum hardware, it is essential to acknowledge the presence of noise in the implemented operations, including those for state preparation and readout.
        In its limiting behavior, oQMLA can be restricted to using only separable states as input and apply solely single-qubit rotations before measurement, which is equivalent to measuring local observables only. 
        We generate a dataset simulating a two-qubit system evolving according to the model
        \begin{equation*}
        \phantom{.}-0.3 \cprim{ZI} - 1.5\cprim{ZX} + 0.2\dprim{\sigma_- I}.
        \end{equation*}
        For each time step we prepare a separable state, simulate the evolution according to the true model, apply a measurement circuit composed of tensor products of single-qubit rotations, and measure in the computational basis. 
        
\begin{figure}[t!]
            \centering
            \includegraphics{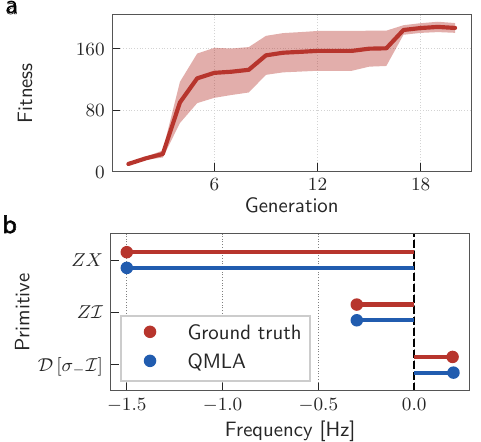}
            {\phantomsubcaption\label{fig: local operations - convergence}}
            {\phantomsubcaption\label{fig: local operations - lollipop}}
            \caption{
                \emph{Local operations} - We simulate the evolution of a two qubits open system and use oQMLA to learn it. 
                Initial states are chosen to be separable, and measurements are of local operators. 
                \textbf{a)}, Evolution of the mean fitness value, computed on the best model of $5$ independent executions. 
                \textbf{b)} Comparison between oQMLA's output and the true model, both in terms of primitives and respective parameters.
            }
            \label{fig: local operations} 
        \end{figure}

        The mean fitness value as a function of the generation number is shown in Fig.~\ref{fig: local operations - convergence}, while Fig.~\ref{fig: local operations - lollipop} illustrates the structure of the learned model compared to the true one.
        Comparing the evolution in Fig.~\ref{fig: local operations - convergence} to the first simulation in this section, we observe that restricting the learning procedure to separable states results in a slower average convergence rate.
        Additionally, despite converging to similar fitness values, the five executions underwent different evolutions, as indicated by the large dispersion in the middle section of the plot.
        This variability is expected for a model like the one we simulate, where the two qubits undergo correlated evolution. 
        In such cases, measuring local operators can provide misleading information, while a basis of entangled states may potentially decouple the outcomes.
        Nevertheless, we see in Fig.~\ref{fig: local operations - lollipop} that the structure of the learned model faithfully resembles the true one. 
        For this reason, in the upcoming simulations, we utilize a dataset where half of the measurements are of local observables.
        This approach aims to facilitate the learning of both independent and correlated evolutions.
        
        \subsection{Noisy data}
        \label{sec: measurement errors}
        In addition to the presence of continuous errors in the state preparation and measurement circuits, another effect that might be relevant in the application of oQMLA is the presence of noise in the measurement outcomes, resulting in their mis\mbox{-}classification. 
        Although it is possible to characterize measurement errors and account for their effects in post-processing~\cite{benjamin_nachman_unfolding_2020}, our objective in this section is to examine whether realistic measurement error rates have a detrimental impact on the learning procedure.
        We prepare a dataset simulating the evolution of a two-qubit system according to the model
        \begin{equation*}
            \phantom{.}1.5\cprim{XZ} + 0.3\dprim{I\sigma_-}+0.2\dprim{Y\sigma_+}.
        \end{equation*}
        For each time step, we simulate the dynamics of an arbitrary pure state and its measurement in an arbitrary basis. 
        The resulting outcomes are then processed through a classical bit\mbox{-}flip channel $\mathcal{N}_{pq}$ which flips $0$s to $1$s with probability $p$ and applies the inverse process with probability $q$. 
        We set the error rates to be on average $2\%$, a realistic value chosen after those reported by IBM for the \verb|ibm_lagos| device~\cite{noauthor_ibm_2021}. 
        
\begin{figure}[t!]
        \centering
        \includegraphics{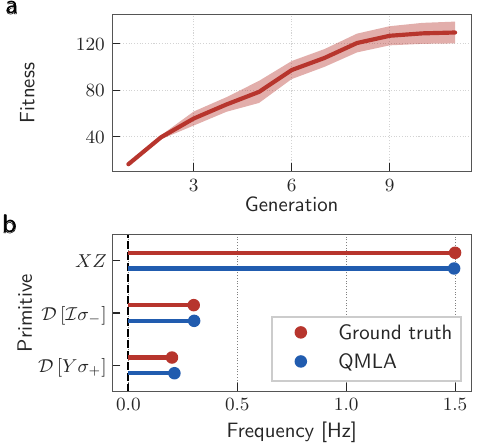}
        {\phantomsubcaption\label{fig: measurement errors - convergence}}
        {\phantomsubcaption\label{fig: measurement errors - lollipop}}
        \cprotect\caption{\emph{Measurement errors} - We simulate the evolution of a two-qubit open system and use oQMLA to learn it. 
        We simulate the presence of measurement errors by stochastically flipping the bits of the measured outcomes. 
        \textbf{a)}, Evolution of the mean fitness value, computed on the best model of $5$ independent executions. 
        \textbf{b)} Comparison between oQMLA's output and the true model, both in terms of primitives and respective parameters.} 
        \label{fig: measurement errors} 
        \end{figure}

        Fig.~\ref{fig: measurement errors} shows the results characterizing the convergence of the algorithm and the output model. 
        The data represented in Fig.~\ref{fig: measurement errors - lollipop} indicate that readout errors with low probabilities hardly have any effect on the parameter estimation routine, and hence on the final output of oQMLA. 
        The reason for this is that we probe the evolution of the system for variable initial states and execution times. 
        Consequently, the effects of an over- and under-estimation of the outcome probabilities average out during the learning process. 
        The most important impact of noisy data is in the value of the fitness function, which converges to mean errors higher than $8\text{\textperthousand}$. 
        This is explained by the fact that the fitness function in Eq.~\eqref{eq: fitness} quantifies deviations from the observed probabilities. The presence of errors in the measured data results in a lower bound to the achievable RMSE, and hence in an upper bound to the value of the fitness function.
        
        \subsection{Approximate-only solutions}
        \label{sec: approximate solution}
        \begin{figure}[ht!]
        \centering
        \includegraphics{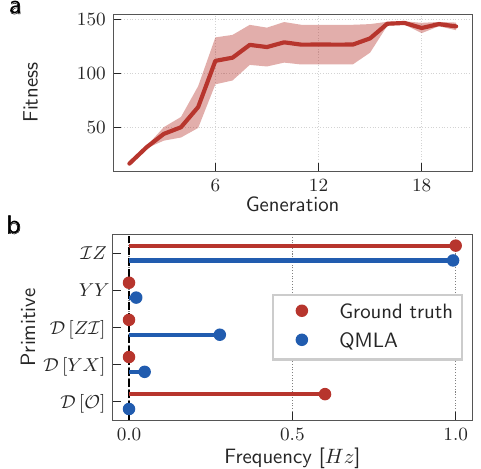}
        {\phantomsubcaption
        \label{fig: approximate solution - convergence}}
        {\phantomsubcaption
        \label{fig: approximate solution - lollipop}}
        \caption{\emph{Approximate-only solution} - We simulate the evolution of a two-qubit open system and use oQMLA to learn it. 
        The jump operator used to simulate the system is not in the set of available primitives $\mathcal{S}$, and hence it cannot be learned. 
        \textbf{a)} Evolution of the mean fitness value, computed on the best model of $5$ independent executions. 
        \textbf{b)} Comparison between oQMLA's output and the true model, both in terms of primitives and respective parameters. 
        The presence of the spurious primitive $YY$ shows the resilience of oQMLA against irrelevant terms. 
        Moreover, oQMLA learned the two jump operators composing the true one, which could not have been found.}
        \label{fig: approximate solution} 
        \end{figure}
        
        Finally, we analyze the behavior of oQMLA when applied to a system whose evolution cannot be learned exactly.
        We generate a dataset simulating the evolution of the model
        \begin{equation*}
            \phantom{.}\cprim{IZ} + 0.6 \dprim{0.3 YX + 0.7 ZI}.
        \end{equation*}
        Note that, although both $YX$ and $ZI$ are in $\mathcal{S}$, the jump operator $\mathcal{O} = 0.3 YX + 0.7 ZI$ is not in the set of available primitives discussed in section~\ref{sec: model}, and hence the true model cannot be found by oQMLA. 
        
        In this example, the output of the algorithm yields a fitness score of around $150$ (cf. Fig.~\ref{fig: approximate solution - convergence}), which corresponds to a mean error of approximately $6.7\text{\textperthousand}$. 
        This indicates that although oQMLA is unable to identify the exact model for the evolution of the system, it can find a good approximation of it.
        The resulting model is reported in Fig.~\ref{fig: approximate solution - lollipop} together with a comparison to the true one. 
        The coherent evolution has been correctly identified, and its rate also closely matches the true one. 
        Regarding the dissipative part, oQMLA identified the two terms composing the true jump operators as the most likely ones.
        Note however that their rates in the learned model are not the same as those in the ground truth.
        This discrepancy is expected because the Bayesian inference sub-routine tries to compensate for the missing cross-terms arising from the evaluation of $L\rho L^\dagger$ and $\acomm{L^\dagger L}{\rho}$. 
        Furthermore, Fig.~\ref{fig: approximate solution - lollipop} highlights an essential property of our fitness function: its resilience towards the presence of unimportant primitives such as $YY$ in this example. 
        Despite employing model reduction to eliminate irrelevant primitives, it is possible for some spurious terms with rates close to $0$ to survive the genetic step and be included in the output model. 
        While this would be desired if the Bayesian inference was able to estimate the parameters to machine precision, allowing oQMLA to capture arbitrarily small effects, it is in this case due to an imperfect training of the model. 
        Hence, it is crucial to employ a figure of merit that is robust against such terms.
        The fitness function in Eq.~\eqref{eq: fitness} accomplishes this as the presence of a small additional term in the model does not significantly alter its evolution and predictive power.


    \section{Hardware implementation}
        The previous section showed the performance of oQMLA in simulated settings.
        Here, we aim to demonstrate the algorithm's practicality on real quantum hardware. 
        Due to their ubiquity, we focus on the noise arising from the application of CNOT gates~\cite{nielsen_gate_2021, barenco_elementary_1995} on IBM Quantum's superconducting computer \verb|ibm_lagos|~\cite{noauthor_ibm_2021}.
        Since the CNOT gate is involutive, applying it twice to the same control and target qubits would, in the ideal case, leave their state unchanged. 
        Nonetheless, due to the imperfect implementation of the CNOTs,  we expect to detect certain discrepancies in the outcome statistics between ideal and experimental realizations. 
        oQMLA can be employed to learn the evolution generated by the noisy CNOT gates, providing insightful information on the system and potentially enabling the application of error mitigation procedures.
        To interface oQMLA with the quantum computer we use the Qiskit Python library~\cite{qiskit_contributors_qiskit_2023, r_wille_ibms_2019}, which allows the execution of quantum circuits both on real hardware and on a classical simulator mimicking the noise of the physical machine.
        
        \begin{figure}[b!]
        \centering
        \includegraphics{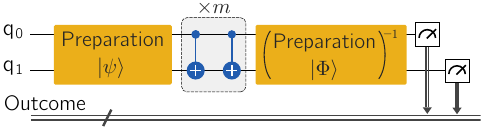}
        \caption{\emph{Quantum circuit defined and executed via Qiskit} - The state preparation gate is used to transform $\ket{00}$ to the input state $\ket{\psi}$ and to map a measurement in a basis containing $\ket{\Phi}$ to one in the computational basis. 
        Between them, $2m$ CNOT gates are applied, with control on the first qubit and target on the second one.}
        \label{fig: qiskit circuit}
        \end{figure}
        
        Following a methodology akin to the previous section, we select an initial state for the two qubits. 
        Subsequently, we apply $m$ pairs of CNOT gates with the first qubit as the control and the second as the target, repeating the procedure for different choices of $m$. 
        Finally, we measure the qubits in a basis containing an arbitrarily chosen state $\ket{\Phi}$. 
        Both the state preparation and measurement circuits make use of the \emph{state preparation gate} provided by Qiskit. In the measurement procedure this process is inverted, mapping a projective measurement onto $\op{\Phi}$ to one onto $\op{00}$. 
        
        Notably, the particle guess heuristic in Eq.~\eqref{eq: particle guess heuristic} relates the experiment time to the uncertainty of the parameters to be estimated.
        For this reason, $t$ can take any floating point value.
        On the other hand, the circuit in Fig.~\ref{fig: qiskit circuit} can only be measured after an even number of CNOTs has been applied. 
        To adapt the particle guess heuristic to the quantum circuit picture, we chose to rescale the time parameter such that in one unit of time $10$ pairs of CNOTs are applied.
        Rounding $t=1/\sigma$ to the first decimal, the particle guess heuristic can be used as is,
        \begin{peq}{,}
            t = \Bigl\lfloor \frac{1}{\sigma}\Bigr\rceil_1 = \frac{m}{10}
        \end{peq}
        where with $\lfloor \cdot \rceil_1$ we indicate that the argument is rounded to the closest first decimal place. 
        Note that in showing the results of the simulations we re-scaled the times and rates in terms of the actual CNOT duration on the first two qubits of the chosen device, which IBM reports to be \SI{576}{\nano\second}~\cite{egger_study_2023}.
        
        \subsection{Proprietary simulator}
        \label{sec: proprietary simulator}
        We initiate our investigation by employing oQMLA on classically simulated data, with the noise model of the \verb|ibm_lagos| processor. 
        We generate a dataset executing the circuit in Fig.~\ref{fig: qiskit circuit} for $m$ ranging from $1$ to $50$. 
        For each value of $m$, we conduct $30$ independent experiments with different initial states $\ket{\psi}$ and probe states $\ket{\Phi}$, repeated $5000$ times each to collect statistical information about the outcomes.
        Notably, it is crucial to emphasize that the true model governing the evolution of the system is unknown for this simulation.
        Therefore, the selection of the target number of primitives is solely based on our intuitive understanding of the potential characteristics of the noise model.
        As argued in section~\ref{sec: reproduction}, sub-optimal choices for this number do not impact the results themselves, but primarily influence the convergence speed of the algorithm. 
        In the absence of additional information regarding the true model describing the system, we opt for a set of seven coherent and dissipative primitives.
        
        \begin{figure}[ht!]
        \centering
        \includegraphics{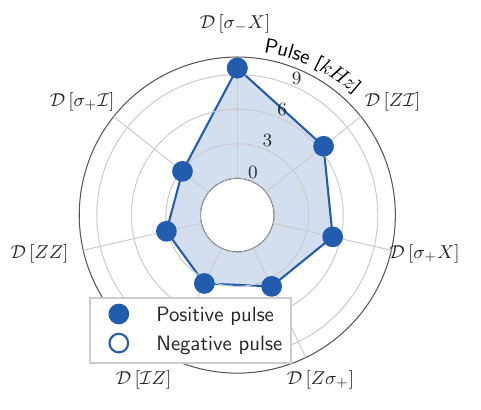}
        \cprotect\caption{\emph{Simulated noise model of CNOT gates } - Output of the oQMLA procedure applied to simulated data. 
        The train- and test-sets have been obtained simulating the circuit in Fig.~\ref{fig: qiskit circuit} with the noise model from \verb|ibm_lagos|. 
        The rays represent different primitives in the model, while the distance from the inner circle is proportional to the corresponding rate.}
        \label{fig: qiskit simulator}
        \end{figure}
        
        A satisfactory approximation of the correct model is already achieved by the fifth generation, enabling predictions of the system's evolution with a mean error below  $2\%$.
        Notably, the maximum score attained by oQMLA stands at approximately $70$.
        This result can be attributed to the increased complexity of the model, as seen in Fig.~\ref{fig: qiskit simulator}, and to measurement errors, reported by IBM to average around $1.5\%$.
        Another notable observation from the model presented in Fig.~\ref{fig: qiskit simulator} is the lack of coherent terms within the learned model.
        Although this result may seem unlikely for the noise model of a real quantum device, it is in fact expected given that the simulator uses incoherent channels to model hardware errors~\cite{noauthor_device_nodate}.
        Nonetheless, results in Fig.~\ref{fig: qiskit simulator} still provide insightful information, as the parameters of such channels are optimized to retrieve the $T_1$ and $T_2$ times measured on the machine~\cite{noauthor_noisemodelfrom_backend_nodate}.
        Consequently, it is reasonable to expect that the real device will exhibit evolution with rates similar to those that we report here, allowing to bias the Bayesian inference procedure by choosing a prior distribution centered around \SI{8}{\kilo\hertz}.
        
        \subsection{Hardware device}
        To obtain more trustworthy information about the noise model of CNOTs on IBM, we repeat the model search with data coming from measurements on the hardware device directly.
        We generate a dataset with the same characteristics as the one we already described in the previous subsection, except for the values of $m$ which we now allow to reach $80$.
        
        \begin{figure*}[ht!]
        \centering
        \includegraphics{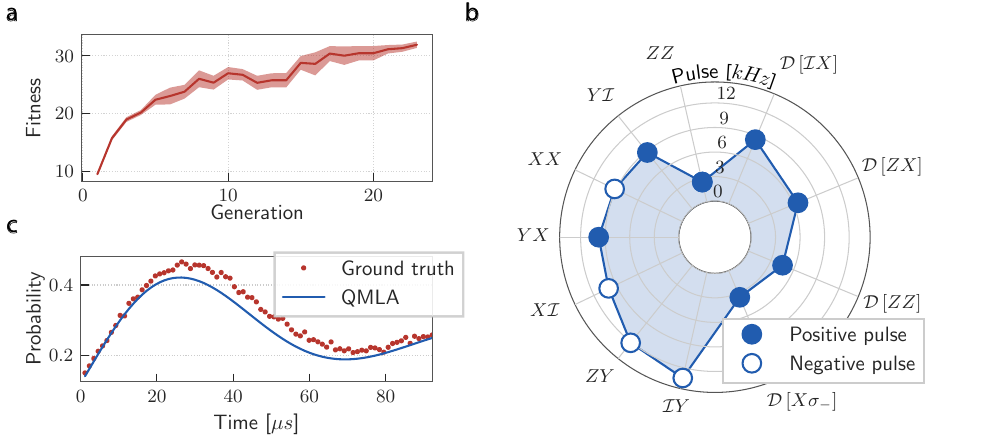}
        {\phantomsubcaption\label{fig: qiskit hardware - convergence}}
        {\phantomsubcaption\label{fig: qiskit hardware - radar}}
        {\phantomsubcaption\label{fig: qiskit hardware - evolution}}
        \cprotect\caption{\emph{Experimental noise model of CNOT gates } - \textbf{a)} Evolution of the mean fitness value, computed on the best model of $5$ independent executions. 
        \textbf{b)} oQMLA's output. 
        The rays represent different primitives in the model, while the distance from the inner circle is proportional to the corresponding rate. 
        For Hamiltonian primitives, an empty marker signals a negative pre-factor. \textbf{c)} Comparison between the dynamic measured on hardware and the one predicted by oQMLA's output.}
        \label{fig: qiskit hardware}
        \end{figure*}
        
        In Fig.~\ref{fig: qiskit hardware - convergence} we report the evolution of the mean fitness function in the number of generations. 
        It is now evident how oQMLA struggles to find a good model for the evolution of the system, with the highest value being around $30$ only. 
        Moreover, the values assumed by the fitness function during the evolution appear to be more variable, with ranges of steady increase followed by local dips~\footnote{Because of this variability, we had to lower the threshold on the fitness value used to determine the convergence of the algorithm to $33$, corresponding to a mean prediction error of approximately $3\%$.}. 
        From the rates of the best model (Fig.~\ref{fig: qiskit hardware - radar}), we deduce that the reason might be its complexity. Bayesian inference, in this case, has to estimate a set of eleven highly correlated rates. 
        Nonetheless, we see from Fig.~\ref{fig: qiskit hardware - evolution} that oQMLA manages to qualitatively predict the dynamics of the system, suggesting that the observed degradation in performance likely arises by two sources.
        On the one hand, measurement errors can significantly reduce the value of the fitness function, as we demonstrated in sections~\ref{sec: measurement errors} and~\ref{sec: proprietary simulator}. 
        On the other hand, as the number of primitives in the model increases, the Bayesian inference routine needs an increasing number of particles to finely resolve the likelihood distribution.
        Each particle requires a full simulation of the dynamics via the Lindblad master equation, resulting in a significant increase of the computational cost.
        Consequently, the number of particles used in the inference must be reduced, which can negatively affect the precision of the parameter estimates.
        Taken together, these considerations suggest that the limited performance is primarily due to the interplay of measurement noise and a poor estimation of the model's parameters, rather than a misidentification of the model’s underlying primitives.


    \section{Discussion and conclusion}
        \label{sec: conclusions}
        In summary, we extended the quantum model learning agent method to the task of characterizing open quantum systems.
        Considering multiple tentative models in a sequential structure, we applied a genetic algorithm to boost the model search. The genetic algorithm looks for the best model combining operators from a pre-defined set. 
        To extend the method to open systems, we included in such a set both Hamiltonian and jump operators and described the evolution through the diagonal form of the Lindblad master equation.
        This choice allows to limit the dimension of the space where the search is performed, effectively speeding up the convergence. 
        At the same time, however, it can affect the quality that the output can reach.
        A possible improvement to the method would be to dynamically add new jump operators to the set of available primitives. 
        Indeed, as we saw in section~\ref{sec: approximate solution}, the presence of two jump operators in the output model can be an indication that the underlying model contains one of their combinations.
        An exploration strategy that, during the model search, analyzes the models in a generation to propose new possible jump operators could be implemented thanks to oQMLA's modular structure. 

        Another advantage of \emph{oQMLA}'s modularity is that it enables the seamless replacement of the simulation backend with alternative computational schemes. 
        This allows, for instance, replacing the classical simulation by evaluations of outcome probabilities on a trusted quantum processor. 
        Similarly, more general classical evolution methods can be implemented, overcoming the limitations of the Lindblad formalism in regimes where its foundational approximations are not valid~\cite{Settineri2018, Zhang2022, Jin2008, milz_2021}.
        
        Another possible advancement to \emph{oQMLA} would be that of using the non-diagonal version of the Lindblad master equation. 
        This choice would allow to restrict the set of available primitives to operators of a complete basis. 
        This will come at the expense of the optimization complexity. 
        Although the genetic algorithm employed in this work achieved promising results in efficiently searching in the model space, it is essential to acknowledge the potential advantages of utilizing collective intelligence-based methods in similar tasks.
        Leveraging a collaborative decision-making process, these methods could offer novel insights and complementary solutions to the complex optimization problem we are tackling, leading to enhanced model exploration and convergence.
        In this regard, we identify as a possible substitute for the genetic algorithm another biologically-inspired method: \emph{Ant colony optimization} (ACO)~\cite{gambardella_solving_1996, dorigo_ant_1996}, previously used in path planning problems.
        An exploration strategy starting from simple models of a single primitive and gradually adding layers of complexity through ACO would allow a fast convergence while ensuring a thorough exploration of the model space.
        Moreover, such greedy searches naturally favor models of a few primitives only, which we identified as more plausible to be correct from a physical standpoint. 

        Defining a fitness function based on the \emph{root mean squared error} allowed for an unbiased evaluation of the performance of models throughout different branches as well as different executions of the algorithm.
        Although the simulations in section~\ref{sec: benchmarking} prove that assessing a model's quality via our figure of merit allows the genetic algorithm to quickly converge to good approximations of the true model, in section~\ref{sec: measurement errors} we observed that measurement errors in the test set lower the accuracy of the characterized model.
        A possible analysis aimed at increasing the performance of oQMLA would be to include a pre-processing phase for the measured outcomes.
        By employing a classical bit-flip channel $\mathcal{N}$, it would be possible to account for the presence of measurement errors both during training and evaluation of a model. 
        This technique would account for any error that can be modeled as a classical channel, provided that it can be prior characterized. 
        
        Finally, we demonstrated that oQMLA can be used to characterize physical systems undergoing an unknown evolution. 
        Here we saw that, even though the output model is qualitatively good, the Bayesian inference fails to provide a robust parametrization when the number of primitives increases.
        This is likely due to the large number of samples needed to faithfully represent the probability function $\Pr(\va{x}|d)$, which has $\qty|\va{x}|$-dimensional domain.
        Analogously to the model search task, finding the best parametrization of a model could also benefit from the adoption of a more advanced algorithm. 
        \emph{Particle swarm} optimization (PS)~\cite{kennedy_particle_1995}, for instance, looks for the best parametrization by evolving multiple agents in the parameter space, allowing them to share information on the explored landscape.
        From the point of view of the dimensionality of the problems, PS has been shown to reliably provide good parametrizations using only a few agents~\cite{piotrowski_population_2020}.
        Alternatively, the use of a differentiable integrator of the Lindblad master equation could be employed, opening the possibility of using gradient-based optimization methods to find the best parametrization of the model~\cite{guilmin2024dynamiqs, alberto_mercurio_2024_14191116}.
        
        In conclusion, we expect our method to provide a robust framework for inferring unknown quantum evolutions in open systems. 
        Through the analysis of its results, we envision oQMLA offering a significant advantage in fine-tuning quantum hardware and the development of platform- and noise-specific error mitigation protocols. 
        \vspace{2mm}


    \section*{Code and Data Availability}
        The data that support the findings of this study are available from the authors upon reasonable request.


    \section*{Acknowledgments}
        The authors thank Brennan de Neeve and Elias Zapusek for helpful comments and discussions throughout the project. 
        This work was supported by the Swiss National Science Foundation (SNSF) through the National Centre of Competence in Research - Quantum Science and Technology (NCCR QSIT) grant 51NF40–160591. 
        I.R. and F.R. acknowledge financial support by the Swiss National Science Foundation (Ambizione grant no. PZ00P2$\_$186040). 
        We acknowledge the use of IBM Quantum services for this work. The views expressed are those of the authors, and do not reflect the official policy or position of IBM or the IBM Quantum team.


    \bibliography{bibliography}
    
\end{document}